\documentclass[fleqn,10pt]{article}

\usepackage{authblk}

\usepackage{color}
\usepackage{float}
\usepackage{booktabs}
\usepackage{units}
\usepackage{amsmath}
\usepackage{amssymb}
\usepackage{graphicx}
\usepackage{esint}
\usepackage{appendix}
\usepackage{times,mathptm}
\usepackage{sectsty}
\usepackage{balance} 
\usepackage[text={18cm,20.4cm},centering]{geometry} 

\usepackage{graphicx} 
\usepackage{lastpage}
\usepackage[format=plain,justification=raggedright,singlelinecheck=false,
font=small,labelfont=bf,labelsep=space]{caption}

\begin{document}

\noindent\LARGE{\textbf{ Pedestrian flows through a narrow
doorway: effect of individual behaviours on the global flow and
microscopic dynamics}}
\vspace{0.6cm}

\noindent\large{\textbf{Alexandre NICOLAS$^{1,2,\star}$ and Sebasti\'an
BOUZAT$^2$ and Marcelo N. KUPERMAN$^2$}}\vspace{0.5cm}

$^1$ LPTMS, CNRS \& Univ. Paris-Sud, F-91405 Orsay, France

$^2$ CONICET \& Centro At\'omico Bariloche, Bariloche, 8400, Argentina

\footnotetext{$^\star$ alexandre.nicolas@polytechnique.edu \newline Phone:
(+33)1.69.15.73.10}

\noindent\textit{\small{\textbf{Received Xth XXXXXXXXXX 20XX, Accepted Xth
XXXXXXXXX 20XX\newline
First published on the web Xth XXXXXXXXXX 200X}}}

\begin{abstract}

We study the dynamics of pedestrian flows through a narrow doorway by
means of controlled experiments.
The influence of the pedestrians' behaviours is investigated by prescribing a
selfish attitude to a fraction $c_s$ of the participants, while the others
behave politely. Thanks to an original setup enabling the re-injection of
egressed participants into the room,  the analysis is conducted in
a (macroscopically) quasi-stationary regime.
We find that, as $c_s$ is increased, the flow rate $J$ rises, interpolating
between published values for egresses in normal conditions and measurements for
competitive evacuations. The dependence of several flow properties on the
pedestrian
density $\rho$ at the door, independently of $c_s$, suggests that
macroscopically the behavioural
aspects could be subsumed under the density, at least in our specific settings
with limited crowd pressure. In particular, under these conditions, $J$ grows
monotonically with $\rho$ up to ``close-packing'' ($\rho \approx
9\,\mathrm{m^{-2}}$). The flow is then characterised microscopically. Among
other quantities,  the time lapses between
successive escapes, the pedestrians' waiting times in front of the door, and
their angles of incidence are analysed statistically. In a nutshell, our main
results show that the flow is orderly for polite crowds, with narrowly
distributed time lapses between egresses,
while for larger $c_s$ the flow gets disorderly and vanishing time lapses
emerge. For all $c_s$, we find an alternation between short and long time
lapses, which we ascribe to a generalised zipper effect. The average waiting
time in the exit zone increases with its occupancy. The disorder in the flow
and the pressure felt by participants are also assessed.

\emph{Keywords: } Pedestrian evacuation; behaviours; exit capacity; bottleneck
flow
\end{abstract}

\newcommand{\comment}[1]{{\textcolor{red}{#1}}}

\thispagestyle{empty}

\section*{Introduction}

Evacuation through a doorway or a narrowing is a long-standing issue. In Homer's
\emph{Odyssey}, Odysseus and his men need to escape from a cave without being
noticed by its blinded guardian, Polyphemus the Cyclops, who relies on his sense
of touch to mount guard; they manage to do so by tying themselves to the
undersides of sheep. Nowadays, the possibility to evacuate quickly and safely
from a public facility or building in emergency conditions should be ensured by
the compliance with building codes
\cite{buchanan2001fire,dinenno2008sfpe,daamen2012emergency}. These codes may be
prescription-based or performance-based \cite{yung1997modelling}, depending on
whether they constrain certain design elements, e.g., door widths, or set
quantitative goals, such as permitting the evacuation of $N$ persons within a
given time. A central issue is then the assessment of exit capacities $J$, that
is, the number of people who can walk through a given exit per unit time. These
exit capacities are not relevant only for emergency evacuations, but more
generally for the dimensioning of pedestrian facilities. Often,
$J$ is evaluated on the
basis of tabulated values resulting from empirical observations or controlled
experiments. The required amount of tabulated data can be reduced by defining a
specific capacity $J_s$, i.e., a capacity per metre width, so that $J= J_s\,w$
for a narrowing of width $w$. This 
simplification relies on the assumption of a roughly linear growth of $J$ with
$w$, which is supported by some experimental
evidence \cite{kretz2006experimental,seyfried2009new,zhang2011transitions}, at
least at moderate pedestrian densities and for $w>60\,\mathrm{cm}$. \footnote{In
fact, among the references supporting an almost constant specific capacity $J_s$
when the exit width $w$ is varied, Kretz et al.~\cite{kretz2006experimental}
reported a moderate decrease of $J_s$ with $w$, whereas Seyfried et
al.~\cite{seyfried2009new} found a slight increase. Reference
\cite{hoogendoorn2005pedestrian} stands out in the literature, in that it
suggests a stepwise increase of the capacity with $w$, due to the formation of
equally spaced lanes, but this hypothesis stems from very limited evidence (only
two widths were studied).}

Unfortunately, as often emphasised in the literature \cite{seyfried2009new}, the
tabulated capacities vary widely between handbooks or papers: for instance, for
constrictions of width $w=1\,\mathrm{m}$, we find $J_s=J=1.30\,\mathrm{s^{-1}}$
in the SFPE handbook \cite{dinenno2008sfpe}, $J=1.60\,\mathrm{s^{-1}}$ in the
``Planning for foot traffic flow in
buildings'' \cite{predtechenskii1978planning}, $J=1.85\,\mathrm{s^{-1}}$ in
Kretz
et al. \cite{kretz2006experimental}, and $J=1.90\,\mathrm{s^{-1}}$ in Seyfried
et
al.~\cite{seyfried2009new}.
These discrepancies are ascribed to various factors. To start with, the use of
distinct measurement methods, e.g., the recourse to either spatial or temporal
averages to define the flow or the density, has been incriminated and can indeed
lead to different results, especially at high densities, even in a simple
single-file flow \cite{seyfried2010enhanced}; but Zhang et al. came to the
conclusion that this point only has a relatively minor impact at lower densities
\cite{zhang2011transitions}. Besides this technical aspect, cultural differences
between populations may play a role; their importance is however debated,
notably owing to the similarity in the measurements performed in the London
Underground (UK) and in Osaka business district (Japan), as mentioned by
Seyfried et al. \cite{seyfried2010enhanced}. The composition of the population
and the associated morphological differences are arguably of greater importance,
with higher flow rates for crowds consisting of children
\cite{daamen2012emergency} ($J_s= 
3.31\,\mathrm{m^{-1}\cdot s^{-1}}$), due to their smaller size. Furthermore, it
was found that short bottlenecks (mimicking the passage through a doorway)
allowed higher flow rates than longer ones, which simulate the entrance to a
narrow corridor \cite{liddle2009experimental,liddle2011microscopic}; in the
former case, it is possible to slither around the door jamb. Perhaps for similar
reasons, trying to predict an exit capacity on the basis of the peak value of
the flow rate \emph{vs.} density curve for a simple uni-directional flow (the so
called fundamental diagram) tends to underestimate the actual capacity
\cite{seyfried2009empirical}. Even more notably, the density of the pedestrian
crowd at the beginning of the egress strongly impacts the flow rate
$J$,
with a faster flow for initially denser crowds
\cite{nagai2006evacuation,seyfried2010enhanced}. Not unrelated to this
dependence on the initial configuration is the question of the stationarity of
the pedestrian flow \cite{hoogendoorn2005pedestrian,
seyfried2009new,liddle2009experimental,
liddle2011microscopic,garcimartin2016flow}: often the measurements are not
performed in the stationary
regime, which further complicates the interpretation of the results.

To study the evacuation dynamics in simpler, or experimentally more tractable,
systems, researchers have investigated constricted flows of less complex
entities, such as vibrated grains \cite{Zuriguel2014clogging},
ants \cite{shiwakoti2011animal,soria2012experimental}, and sheep
\cite{garcimartin2015flow,zuriguel2016effect}, and brought to light similarities
with pedestrians (for ants, the analogy is however debated
\cite{Parisi2015faster}). But the complex psychology of humans adds a level of
complexity to their response. In particular, on account of their individual 
characters, pedestrians exhibit diverse reactions to external stimuli and
variable behaviours. In a series of repetitive evacuation experiments,
habituation and boredom may lead to considerable variations between repetitions
\cite{daamen2012emergency}. Besides, most of the aforementioned research
focuses on pedestrian flows in normal, cooperative conditions,
while competitive
evacuation experiments are rare, probably because of the risks they present,
although in 
reality impatient, competitive, aggressive or vying behaviours are also observed
in some situations
\cite{casanova1863resumen,hatch2003tinder,kugihara2001effects,
nicolas2016statistical}. Among but a few other examples, Muir et
al. \cite{muir1996effects} simulated the evacuation from an aircraft, in which
the participants' competitiveness was whetted by the possibility of a reward;
Nagai et al. \cite{nagai2006evacuation} investigated the evacuation of
relatively competitive crawlers and walkers;  Helbing et al. conducted
panic-like evacuation experiments, but with only about 20 participants. 
(Also see Ref.~\cite{hoogendoorn2003extracting,daamen2003experimental} for a
study of uni-directional flows, with some prescribed participants' behaviours). 
More recently, in a series
of papers \cite{Garcimartin2014experimental,Pastor2015experimental,
garcimartin2016flow}, Garcimart\'in, Zuriguel, and co-workers carried out a
detailed analysis of controlled evacuations in which distinct levels of
competitiveness were prescribed to the crowd. (Also see \cite{moussaid2016crowd}
for experiments in a virtual environment.) It is noteworthy that, 
compared to the values reported above for normal conditions, the specific
capacities measured in those (diversely) competitive settings are considerably
larger: $J_s=3.3\,\mathrm{m^{-1}\cdot s^{-1}}$ in the work of Nagai et
al. \cite{nagai2006evacuation} and Helbing et al. \cite{helbing2005self},
$J_s\approx 3.7\,\mathrm{m^{-1}\cdot s^{-1}}$ in Garcimart\'in et al.'s
paper \cite{garcimartin2016flow} for door widths of 70 to 80 cm.

However, these evacuations were performed with crowds of homogeneous
competitiveness. To get greater insight into the effect of individual
behaviours,  we conduct and analyse controlled evacuations through
a narrow doorway with participants that are prescribed \emph{distinct}
behaviours, namely, either a polite or a selfish behaviour. We also vary their
eagerness to egress. Note that
`evacuation' will be employed in the broad sense of emptying of a room with no
focus on emergency conditions. For obvious safety reasons, our controlled
experiments do not include many of the aspects that may occur in situations of
extreme emergency such as panic, high pressure, violence or extreme haste. In
contrast, they actually correspond to the flow of a dense crowd 
through a narrow exit with variable eagerness to exit, but always limited
pushing. Although the behavioural
aspects add yet another component to a complex problem, the global picture of
the evacuation process is
clarified by considering a (well-nigh) stationary regime and seeking  robust
``microscopic'' characteristics of the flow. This yields a quantitative
characterisation of the dynamics at the bottleneck.

The next section is dedicated to the description of the experiments. We
then analyse the experimental results in terms of global flow
properties, before turning to a more microscopic study. Let us mention that a
concise
account of the \emph{global} flow properties observed in these experiments was
proposed in a preliminary report \cite{nicolas2016influence}.

\section*{Presentation of the evacuation experiments and methods}

\subsection*{Experimental setup}

\begin{figure}[ht]
\centering
\includegraphics[width=400pt]{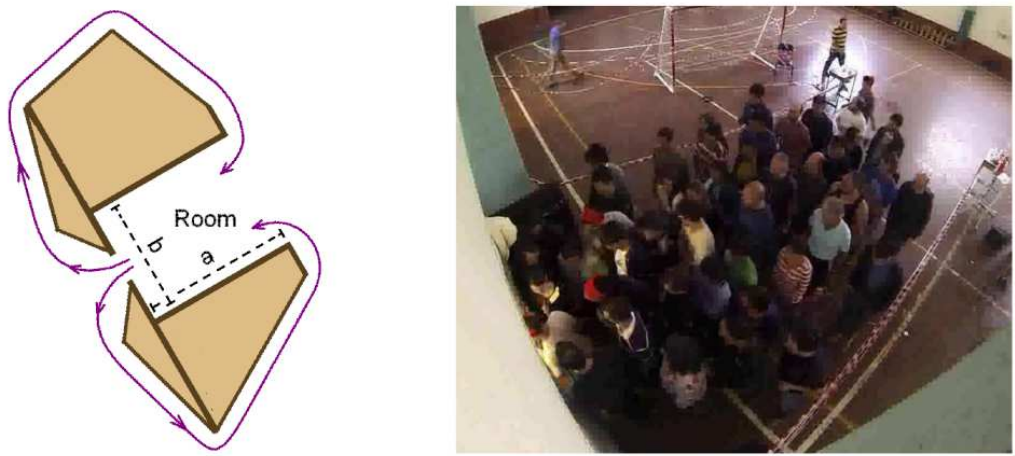}
\caption{Sketch of the experimental geometry (left), with a representation of
the two (short and long) re-injection circuits, and snapshot of one of the
evacuations (right). The dimensions are: $a=7.3\,\mathrm{m}$,
$b=3.5\,\mathrm{m}$, and the door is $72\,\mathrm{cm}$-wide. The short and long
re-injection circuits are approximately $17\,\mathrm{m}$ and $25\,\mathrm{m}$
long, respectively.}
\label{fig:sketch}
\end{figure}

The bottleneck flow experiments were performed in the gymnasium of
Centro At\'omico
Bariloche (CAB), Argentina, and involved
more than 80 voluntary participants (students and researchers), aged 20 to 55
for the greatest part, with a woman/man ratio of about 1:3. The participants
were asked to evacuate a delimited area through a $72\,\mathrm{cm}$-wide
doorway;
the geometry is sketched in Fig.~\ref{fig:sketch}. This doorway was created by
moving $\approx 10 \mathrm{cm}$-thick sliding walls. Safety was a central
concern. Accordingly, the
door jambs were padded with training mats.
In addition, because of the possible ethics and safety issues, the protocol was
validated beforehand by a local ethics committee and
the experiments were prepared in collaboration with the Safety and Hygiene group
of CAB. The evacuation process was supervised
by three staff members who could stop it at any moment by blowing a whistle.

The experimental design is directly inspired by the experiments of Zuriguel's
group \cite{Pastor2015experimental},
but we introduced some important changes. In each evacuation experiment, a
fraction $c_{s}$ of the participants was told to behave selfishly while the rest
should behave politely. The selfish agents were selected randomly and varied
between the experiments; they were asked to wear a red headscarf to be
recognisable on the videos. They were allowed to ``elbow their way through  the
crowd, with mild contacts
but no violence whatsoever''. Their polite counterparts, on the other
hand, were to ``avoid any contact and try to keep their distance''. Before each
experiment, all participants were asked to walk as close as possible
to the door, without crossing it. Only then did we announce who would
behave selfishly in the upcoming run. 

The session started with a mock evacuation. Then, in a first series of
experiments, here referred to as experiments with
\emph{placid} walkers, all participants were instructed to ``head for the
door'', without further specification. In the second
series, they were told to hurry a bit more (\emph{andar con m\'as ganas} in
Spanish), but without
running, pushing or hitting others''; in this case, the walkers are said to be
\emph{in a hurry} or
\emph{hurried}. Therefore, the experiments are controlled by two
orthogonal 'behavioural' parameters, the fraction $c_s$ of selfish participants
and the global placidity \emph{vs.} hurry of the crowd. Overall, the
experiments lasted for somewhat more than
one hour.

Finally, the often reported lack of stationarity in the flow, along with the
word of caution in Ref.~\cite{garcimartin2016flow} about the risk of faulty
statistics with too small crowds, raised strong concerns in our minds about
finite-size effects. To curb these effects, we decided to innovate by imposing
``periodic boundary conditions'' on the crowd:
after egressing, participants followed a circuit that led to their re-injection
into the room. In fact, to optimise the randomisation and limit the clustering
of, say, fast participants, two re-injection circuits
were set up, a short one and a longer one, and the evacuees were directed to
either one alternatively. With this contrivance, about 250 passages through the
door (between 177 and 352, to be precise) were obtained in each experiment.

\subsection*{Observation of uncontrolled egresses}
To complement our controlled experiments, we filmed the egresses from a
conference room at the various breaks of the TREFEMAC conference, a three-day
Physics workshop held at CAB. The participants were informed of the recording
once and for all. Some tendency to align before passing through the 82-cm-wide
exit door was observed; the door was opened by an angle of more than $90^\circ$.
The data,
presented in Fig.~3e, consist of about 260 individual
passages. 
We also filmed two collective egresses from the auditorium of CAB at the end of
the weekly seminars. The 75-cm-wide door was fully open and the 50 to 100
participants were unaware of the recording. Some were carrying a chair when
passing through the doorway. The data, presented in Fig.~3f,
consist of about 150 individual passages.

\subsection*{Video analysis}
The controlled evacuations were recorded with two 60Hz, large-angle Go Pro
cameras and one standard 60Hz camera. Two cameras were placed above the door,
and one filmed the crowd in the room from a more distant point of view. Some of
the videos are available as Supplemental Materials.

\begin{figure}[h]
\centering
\includegraphics[width=220pt]{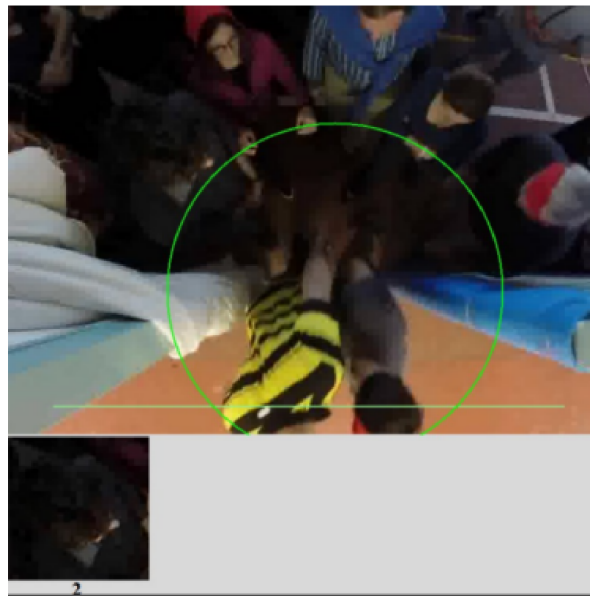}
\caption{One randomly selected video frame, with the exit zone marked in light
green. A home-made Python routine allows us to click on the pedestrians when
they enter the zone and deselect them when they exit.}
\label{fig:snapshot}
\end{figure}

With respect to the image analysis, a standard method to get the egress times is
to build a time line of passages (see Supplementary Fig.~S1) by
extracting a few lines of pixels just past the door from all video frames and
then stitching them together. But to get more comprehensive information, we
resort to a more exhaustive semi-manual analysis. We flick through all video
frames, down-rated to 30Hz,
clicking on the participants' heads when they enter the semi-circular exit
zone shown in Fig.~\ref{fig:snapshot} and pressing a
dedicated key when they cross a line beyond the door. Both methods agree
very
well regarding the exit times.

The density $\rho$ near the door is assessed on the basis of the
occupancy of the exit zone, which is evaluated by detecting the pedestrians'
first
entrance into this zone and their time of egress. The area of this
zone ($0.42\,\mathrm{m^2}$) was chosen small enough to provide an estimate the
local density at the door,
while being large enough to limit the noise in the data.
(Preliminary tests with a slightly larger rectangular zone yield
similar results.) This method creates a small artefact in the most
competitive evacuations, in which a few participants were shortly pushed out of
the zone, due to a crowd movement, before moving in again and passing the door,
a possibility that is discarded in our routine, hence a slight overestimation of
the density. Moreover, no correction for optical distortion was performed,
because we used
the central part of the images. Despite this central position, the parallax
resulting from the camera not being strictly at the zenith (owing to the
impossibility to place it high enough) implies that the moment of detection of a
given participant's entrance into the exit zone somewhat depends on their
height. This results in some inaccuracy with respect to the effective area
of the zone in front of the door. To reflect this uncertainty, we
introduced a pre-factor $\alpha \simeq
1$ in the definition of the density unit ($\alpha\,\mathrm{m^{-2}}$). 
Coarse geometric considerations lead to an upper bound of 30\%, roughly
speaking, on the error in the absolute value of the density in the worst case,
for an uncertainty about the pedestrians' heights of 30cm; this error is however
expected to become smaller upon averaging density over time. All
in all, our density evaluation method has its shortcomings, but given
 the experimental conditions under consideration and our video
recordings, we deem it satisfactory. 

\subsection*{Error bars and confidence intervals}
Wherever relevant, we show error bars on the graphs to give an idea of the
uncertainty due to the finite number of measurements we performed. In these
cases, the data points are obtained as averages over a sub-sample of our
measurements; the error bars are systematically $\frac{2\sigma}{\sqrt{n}}$ on
each side of the plotted data point, where $\sigma$ is the standard deviation of
the sub-sample and $n$ is the sub-sample size.
If the sample points are uncorrelated, the central limit theorem
implies that these error bars represent 95\% confidence intervals. But two
caveats should be heeded. First, not all the observables for which we plot
error bars are uncorrelated. Still, their correlations are short-termed and
weak, with relative magnitudes always below 13\%, so it is reasonable to
neglect these correlations. Secondly, perhaps more importantly,  this
definition of the
confidence interval is valid only in the limit of large $n$, whereas in our case
$n$ is generally of order 100-300 (except for
Fig.~\ref{fig:T_w}(right)).

\section*{Results}

\subsection*{General observations}

The experiments went as planned, without incident. The re-injection process
operated properly, insofar as no gap in the inflow of
pedestrians at the door was seen until the end of each evacuation: The flow
was thus limited by the passage through the doorway, and not by travel times in
the
re-injection circuits.

Some selfish agents tried to overtake the rest of the crowd by walking near the
lateral boundaries of the delimited area, instead of going into the thick of the
crowd. More generally, we observed the formation of files of a few selfish
agents following each other and taking advantage of the `voids' opened by their
predecessors. We believe that this follows from a general mechanism of
clustering via a coupling between the structure of the medium and the motion of
the particles (for instance, the pairing of electrons in a superconductor
\cite{bardeen1957theory}), although it is true that, here, the formation of
files may have been facilitated by the conspicuousness of the headscarf-wearing
selfish agents. Importantly, given that selfish agents made their way faster,
they were re-injected into the room more frequently. Accordingly, the effective
fraction $c_s^\star$ of selfish evacuees is generally larger than the prescribed
(nominal) fraction $c_s$. The results below will be presented as a function of
$c_s^\star$, where relevant, 
rather than $c_s$.

For polite crowds, i.e., at low $c_s^\star$, the egress looks orderly
and
contacts between pedestrians are avoided. Some polite participants yield
deliberately and may even wave to their neighbours to go ahead. Very generally,
there is no more than one pedestrian crossing the door at each time. For larger
$c_s^\star$, the crowd in front of the door becomes more compact, contacts
surge, and the flow looks more chaotic as several pedestrians try to walk
through the door simultaneously. Such endeavours sometimes lead to clogs due to
the formation of `arches' of three to four participants across the door,
particularly at $c_s^\star \geq 90\%$. But these clogs are rapidly resolved, in
about one second or less, and do not \emph{substantially} delay the flow. In the
most competitive evacuations ($c_s^\star=92\%$(hurried) and
$c_s^\star=100\%$(hurried)), some agents grasp the door jamb and use it
to
pull themselves past the door; a few pedestrians spin around upon egressing,
because of the contacts with 
their neighbours.

\subsection*{Macroscopically stationary, but intermittent dynamics}
\begin{figure}[ht]
\centering
\includegraphics[width=\linewidth]{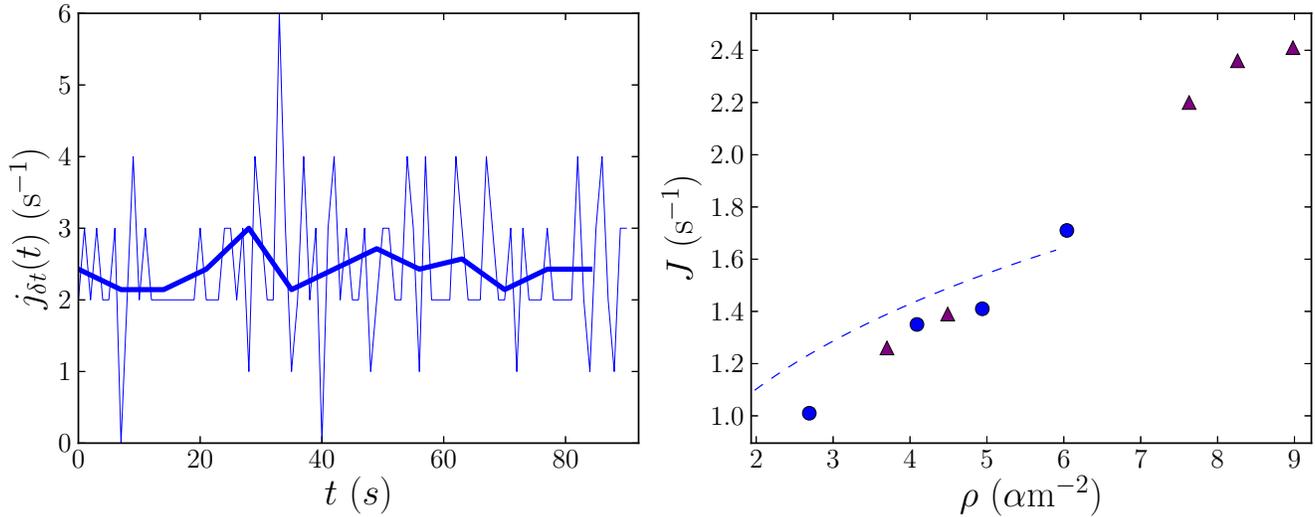}
\caption{(Left) Time-dependent flow rate $j_{\delta t}(t)$ in the evacuation
with $c_s^\star=100\%$ (hurried), for $\delta t= 1\,\mathrm{s}$ (thin
lines)
and $\delta t= 7\,\mathrm{s}$ (thick lines). (Right) Global flow rate $J$  as a
function of the average local density $\rho$ at the door (the corresponding
$c_s^\star$ can be found in Table~1). The dots represent the
experiments with placid participants and the triangles, the hurried
ones. The
dashed line is the prediction of the naive model presented in the text. The
geometrical prefactor $\alpha\simeq 1$ in the density unit is due to
experimental uncertainty.}
\label{fig:flowrates_7}
\end{figure}

To go beyond these general observations, we extract the exit times
$t^\mathrm{out}_i$ of each pedestrian $i$, numbered according to their order of
egress, regardless of their identity (see the \emph{Video Analysis} section for
details). This allows us
to compute the time-dependent flow rate $j_{\delta t}(t)$, which is a moving
average over a time interval $\delta t$, for each evacuation, viz.,  
\begin{equation*}
j_{\delta t}(t) \equiv \frac{1}{\delta t} \sum_i
\Theta(t^\mathrm{out}_i-t)\,\Theta(t+\delta t - t^\mathrm{out}_i),
\end{equation*}
where $\Theta$ is the Heaviside function, viz., $\Theta(x)=1$ if $x\geqslant 0$,
0 otherwise. For  $\delta t=1\,\mathrm{s}$, the curve $j_{\delta t}(t)$ is
extremely jagged, as can be seen in Fig.~\ref{fig:flowrates_7}(left). It
consists of somewhat irregular spikes that are all the larger as the evacuation
is competitive (at large $c_s^\star$). But averaging over a larger time window,
$\delta t=7\,\mathrm{s}$, smears out these spikes (i.e., high-frequency
oscillations) and yields a smoother
curve, whose relative flatness
proves that, macroscopically, the pedestrian flow is quasi-stationary:
there is no upward or downward trend over extended periods of time
(compared to the experiment duration). 
Interestingly, this quasi-stationary regime is reached almost immediately after
the
beginning of the experiments and persists until at least a dozen seconds before
their ends\footnote{The experiment at $c_s^\star = 18\%$(hurried) is a
slight
exception, insofar as the flow rate wanes moderately in the last 30 or 40
seconds.}. The density of pedestrians in the exit zone is also roughly
stationary in general (see Supplementary Fig.~S2).

Our observation of macroscopically quasi-stationary dynamics contrasts with the
non-stationary flows found in a number of related
works \cite{seyfried2009new,liddle2011microscopic,garcimartin2016flow}. We see
two main reasons for this contrast: a 3-m-large gap  (1-m-large in
Ref.~\cite{garcimartin2016flow}) in front of the exit was initially left free of
participants in Ref.~\cite{seyfried2009new}, which explains the existence of a
transient regime, and no re-injection of the participants was enforced in those
earlier experiments.
No matter how limited the transient effects are in our case, to further reduce
them, we henceforth discard the first three and the last ten seconds of all
experiments.

\begin{table}[ht]
\centering
\begin{tabular}{|c|c|c|c|c|}
\toprule
$c_s$ & $c_s^\star$ &  Density $\rho$ ($\alpha \mathrm{m^{-2}}$) & Flow rate $J$
($\mathrm{s}^{-1}$)  & Number of full turns \\
\hline
\hline
\multicolumn{5}{l}{Placid walk: "Head for the door"} \\
\hline
0\%	& 0\%  		& 	2.69 &	1.01 &	0 \\
30\% &	45\%	&  4.09	 &  1.35 &	0 \\
30\% &	47\%	& 	4.94 &	1.41 &	0 \\
60\% &	71\%	&	6.04 &	1.71 &	0 \\
\hline
\multicolumn{5}{l}{Hurried walk: "Head for the door
\emph{more hurriedly}"} \\
\hline
0\% &	0\%		&	3.70 &	1.26 &	0 \\
10\% &	18\%	&	4.49 &	1.39 &	0 \\
60\% &	71\%	&	7.63 & 	2.20 &  3 \\
90\% &	92\%	&	8.26 &	2.36 &	3 \\
100\% &	100\%	&	8.98 &	2.41 &	5 \\
\bottomrule
\end{tabular}
\caption{\label{tab:global}Parameters of the controlled experiments and
measurements: nominal fraction of selfish agents $c_s$, effective selfish
fraction $c_s^\star$, average flow rate (or exit capacity) $J$, average density
$\rho$ in the exit zone, and number of participants spinning around a full
($360^\circ$) turn upon egressing. The experiments were performed in the
following order (by line number): 1-2-4-3-7-8-9-6-5, each line representing one
distinct realisation.}
\end{table}

\subsection*{Dependence of the flow rate on the density}
We now compute the global flow rates $J$ for each experiment as time
averages of the $j_{\delta t}$. The result is independent of $\delta t$
and can also be expressed as $\frac{N-1}{t^\mathrm{out}_N - t^\mathrm{out}_1}$,
where $N$
is the number of passages and $t^\mathrm{out}_1$ and $t^\mathrm{out}_N$ are
the first and last egress times, respectively. The values are listed in
Table~\ref{tab:global}. Overall, $J$ grows monotonically with the concentration
$c_s^\star$ of selfish agents, and a peak value of 2.4 persons per second is
reached at $c_s^\star=100\%$. But $J$ also increases when the placid crowd is
asked to walk with more hurry (irrespective of the individual
behaviours):
for the
same value of $c_s^\star$, namely 71\%, $J$ surges by almost 30\% following this
change. The sensitivity of $J$ to the participants' attitudes, while the
geometry and
the crowd are kept fixed, renders the divergence of published values for the
exit capacity quite understandable. 

Turning to a comparison with these published values,  we see that the specific
capacity corresponding to the placid polite crowd ($J_s =
1.4\,\mathrm{m^{-1}\cdot s^{-1}}$) is comparable to the rather conservative
specification of the SFPE handbook \cite{dinenno2008sfpe}, namely,  $J_s =
1.30\,\mathrm{m^{-1}\cdot s^{-1}}$, while the specific capacity for the polite
but more hurried crowd ($J_s \simeq 1.75\,\mathrm{m^{-1}\cdot s^{-1}}$)
is
similar to Kretz et al.'s measurement \cite{kretz2006experimental} ($J_s \simeq
1.74\,\mathrm{m^{-1}\cdot s^{-1}}$) for a flow through $70\,\mathrm{cm}$-wide,
$40\,\mathrm{cm}$-long bottleneck in normal conditions, and comparable to the
value $J_s \simeq 1.61\,\mathrm{m^{-1}\cdot s^{-1}}$ reported by Seyfried et
al. \cite{seyfried2009new} for a flow through $80\,\mathrm{cm}$-wide,
$2.8\,\mathrm{m}$-long bottleneck in normal conditions. At the other extreme,
the highest flow rate that we measured, $J=2.41\,\mathrm{s}^{-1}$ for
$c_s^\star=100\%$(hurried), is slightly below, but 
similar to, the range of flow rates obtained in the more or less competitive
evacuations of Pastor et al. \cite{Pastor2015experimental} ($J=2.43\text{ to
}2.63\,\mathrm{s}^{-1}$) for a 69-cm-wide (instead of 72 here) door.

The increase of  $J$ with $c_s^\star$ and with the agents' degree of
hurry in our
experiments points to the absence of a `faster-is-slower' effect, or, better
said,
a `more-competitive-is-slower' effect. This may seem to be at odds with the
conclusions of Pastor et al.~\cite{Pastor2015experimental,garcimartin2016flow},
but we claim that there is no contradiction: Unlike
\cite{Pastor2015experimental,garcimartin2016flow}, our experiments are not (all)
above the threshold of competitiveness required to observe this effect. Indeed,
in \cite{Pastor2015experimental,garcimartin2016flow}, competitive participants
were allowed to push each other with stretched arms, unlike in ours; the larger
competitiveness is also reflected by the larger flow rates that they measured.
The granular analogy then helps clarify the consequence of this difference
\cite{Zuriguel2014clogging} : In a granular hopper flow, as the panel supporting
the grains is tilted and the effective gravity increases accordingly, in a first
regime the flow 
becomes faster, because grains fall faster, but then it slows down under higher
gravity. This slowdown is due to the stabilisation of clog-inducing arches by
pressure. Models for pedestrian dynamics premised on Newton's equations with
``social forces'' predict a similar non-monotonic effect of the pedestrians'
desired velocity  \cite{parisi2007faster}. As this velocity is increased, in a
first regime the evacuation gets faster, but a second regime then emerges, where
the repulsive forces between clustered pedestrians at the exit exceed the
driving force associated with the desired velocity; the ensuing clogs
reduce the flow rate. We presume that the evacuations of
\cite{Pastor2015experimental,garcimartin2016flow} are all within the second
regime, whereas ours are not.

\subsection*{Dependence of the flow rate on the pedestrian density}

We have already noted that the flow rate is not uniquely determined by
$c_s^\star$ in our experiments. Following Seyfried et al.'s
\cite{seyfried2009new} idea that the discrepancies between the exit capacities
in the literature are largely due to different initial crowd densities, we
investigate the dependence of $J$ on the average density $\rho$ in the exit zone
(shown in Fig.~\ref{fig:snapshot}). Recall that  the pre-factor $\alpha
\simeq 1$ in the density unit
($\alpha\,\mathrm{m}^{-2}$) was introduced so as to reflect the
experimental
uncertainty in our measurements, as explained in the \emph{Video Analysis} 
section. In any event, the absolute
density values would certainly be different for a different crowd (composed
of, say, bulkier participants), but the evolution of flow properties with the
density is expected to be robust. It is also worth mentioning that the highest
values that we report for the density ($\rho \approx 9 \mathrm{m}^{-2}$) are
obtained for a tightly-packed crowd. These values may appear intriguingly
large; it
should however be emphasised that here $\rho$ is not the average density in the
crowd, but its \emph{local} value near the exit, where pedestrians are most
densely packed; previous works have already shed
light on the considerable deviations between the average density and its peak
local value \cite{helbing2007dynamics,liddle2011microscopic}.

The graph $J(\rho)$ plotted in Fig.~\ref{fig:flowrates_7}(right) confirms the
relevance of the density parameter: the flow rates measured for different
behavioural prescriptions collapse onto a smooth master curve (this is
not achieved when plotting $J$ as a function of $c_s^\star$). The monotonic
increase of the flow rate $J$ with $\rho$ is remarkable.
At low densities, the order of magnitude of $J$ and its growth with $\rho$ can
be rationalised rather straightforwardly. First, notice that pedestrians in the
exit zone are separated by $l=\rho^\frac{-1}{2}$ on average. Besides, they
generally egress one by one and lower their velocity when the door is blocked by
somebody else's passage. Should one crudely assume that this produces a halt,
after which they gather speed as $v(t)= v_0\,\mathrm{min}(\frac{t}{\tau},1)$ 
with a final velocity $v_0\approx 1\,\mathrm{m\cdot s^{-1}}$ and a response time
$\tau \approx 0.4\,\mathrm{s}$, one will easily obtain the time interval
required to walk a distance $l$, $\Delta t=\frac{\tau}{2}+\frac{l}{v_0}$
(provided that $\Delta t>\tau$), hence a flow rate $J=\Delta t^{-1}=\frac{2 v_0
\sqrt{\rho}}{2 + v_0 \tau \sqrt{\rho} }$. This is undoubtedly a very crude
model, but its prediction, represented as dashed lines in
Fig.~\ref{fig:flowrates_7}(right)), is in broad agreement with the low-density
data.

Much more surprisingly, $J$ keeps increasing up to high densities, $\rho\approx
9 \alpha\mathrm{m}^{-2}$, despite the fact that most fundamental
diagrams in the literature suggest a decline of the flow above much smaller
densities, due to jams
\cite{predtechenskii1978planning,weidmann1993transporttechnik,
dinenno2008sfpe,flotterod2015bidirectional} (see Fig.~11 of
Ref.~\cite{seyfried2009new}). The uncertainty in our assessment of the
density can hardly be responsible for this discrepancy: {The tight packing of
the crowd in our most competitive evacuation is compatible with the value
$\rho\approx 9 \mathrm{m}^{-2}$ that we report}.
Instead, at such high densities the flow rate is probably not uniquely
determined by $\rho$, but strongly depends on other parameters, such as the
pressure in the crowd. Indeed, since the tightly-packed crowd is nearly
incompressible, mechanical pressure can increase vastly, hence
stabilising clogs, while $\rho$ changes only little. Crowd turbulence
\cite{helbing2007dynamics,yu2007modeling} may also arise, thereby affecting the 
density and flow. In our settings, excessive mechanical pressures and turbulence
were warded off thanks to the strict instructions given to the participants.

Importantly, the dependence of the flow rate on the density holds not only for
the mean values of these quantities, but often also for their temporal
fluctuations $\delta j_{\delta t}(t)\equiv j_{\delta t}(t) - J$ and $\delta
\rho_{\delta t}(t)\equiv \rho_{\delta t}(t) - \langle \rho \rangle$. Here, the
time-dependent local density $\rho_{\delta t}(t)$ is defined as the average
number of pedestrians in the exit zone between $t$ and $t+\delta t$, divided by
the zone area. Indeed, the correlator
\begin{equation*}
C_{\rho j}^{\delta t} \equiv \frac{\langle \delta\rho_{\delta t}(t) \delta
j_{\delta t} (t)\rangle}{\sqrt{\langle \delta \rho_{\delta t}(t)^2 \rangle
\langle \delta j_{\delta t}(t)^2 \rangle}},
\end{equation*}
where the brackets denote a time average, indicates the existence of
statistically significant positive correlations between the flow rate and the
density fluctuations (Supplementary Fig.~S3), when they are averaged over time
intervals
$\delta t$ of a few seconds. However, correlations are only moderate. It should
be
noted that these correlations are not visible in the most competitive
experiments, $c_s^\star=71\%,\,92\%$ and to a lesser extent 100\%
(hurried).

\subsection*{Distribution of time lapses between successive egresses}

\begin{figure}[ht]
\centering
\includegraphics[width=\linewidth]{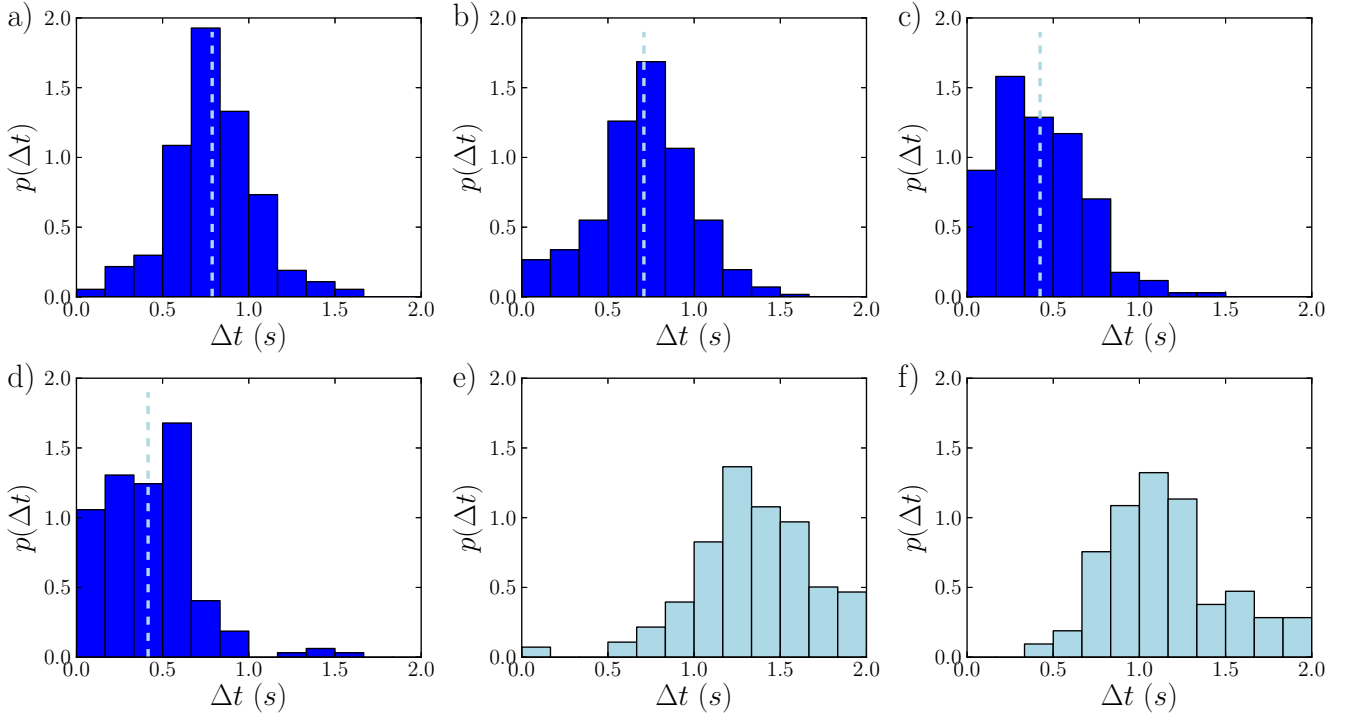}
\caption{Normalised histograms of time lapses $\Delta t$ between egresses for
different evacuations. The first four panels correspond to experiments for (a)
$c_s^\star=0\%$(hurried), (b) $c_s^\star=30\%$(placid), (c)
$c_s^\star=92\%$(hurried), (d) $c_s^\star=100\%$(hurried).
The dashed lines indicate the mean time lapse $\langle \Delta t \rangle$
in the experiment. The
last two
panels correspond to uncontrolled egresses (see the text) from (e) a conference
room during a three-day congress and (f) the auditorium of CAB at the end of the
weekly seminars.}
\label{fig:time_lapses}
\end{figure}

On average, the flow rate is related to the mean time lapse $\langle \Delta t
\rangle$ between successive egresses via $J=\langle \Delta t\rangle^{-1}$. But,
since fluctuations are large, it is actually worth considering the full
distribution $p(\Delta t)$ of time lapses. Figure~\ref{fig:time_lapses} reveals
substantial differences in these distributions when the crowd behaviour is
varied. Pedestrian flows with a mostly polite crowd display a
relatively sharp peak
at a characteristic time $\Delta t^\star$ slightly below one second, and a
depleted region at low $\Delta t$. Interestingly, these features are also
observed in uncontrolled collective egresses from conference rooms through
doorways of similar widths, as shown in Fig.~\ref{fig:time_lapses}e-f (these
uncontrolled egresses were described above). In these uncontrolled
settings, the part of the distribution corresponding to large $\Delta t$ is
irrelevant, insofar as it is due to the foot-dragging of participants
within the conference room, rather than to the dynamics at the exit.

On the other hand, competitive egresses at large $c_s^\star$ (in our controlled
experiments) do not present a very well defined peak and the most frequent time
lapses are shifted to smaller values. Most conspicuously, they include a
non-negligible fraction of very short time lapses $\Delta t\rightarrow 0$, which
correspond to quasi-simultaneous egresses.

Focusing on the large values of $\Delta t$, caused by jams, the data
corresponding to our most competitive evacuation ($c_s^\star=100\%$,
hurried)
seem to support Garcimartin et
al.'s \cite{Garcimartin2014experimental,Pastor2015experimental,
garcimartin2016flow} claim of a slow decay of $p(\Delta t)$ at large $\Delta t$,
characterised by a power-law-like tail: We find a higher likelihood for the tail
to be power-law-like than exponential ($p=0.04$ for a continuous data set,
according to the method of Clauset et al. \cite{clauset2009power} implemented in
\cite{alstott2014powerlaw}). In other situations, our data are insufficient to
validate this claim, and even tend to contradict it in the presence of
polite crowds.

\subsection*{Temporal correlations in the flow}
\subsubsection*{Bursts of escapes}
\begin{figure}[ht]
\centering
\includegraphics[width=290pt]{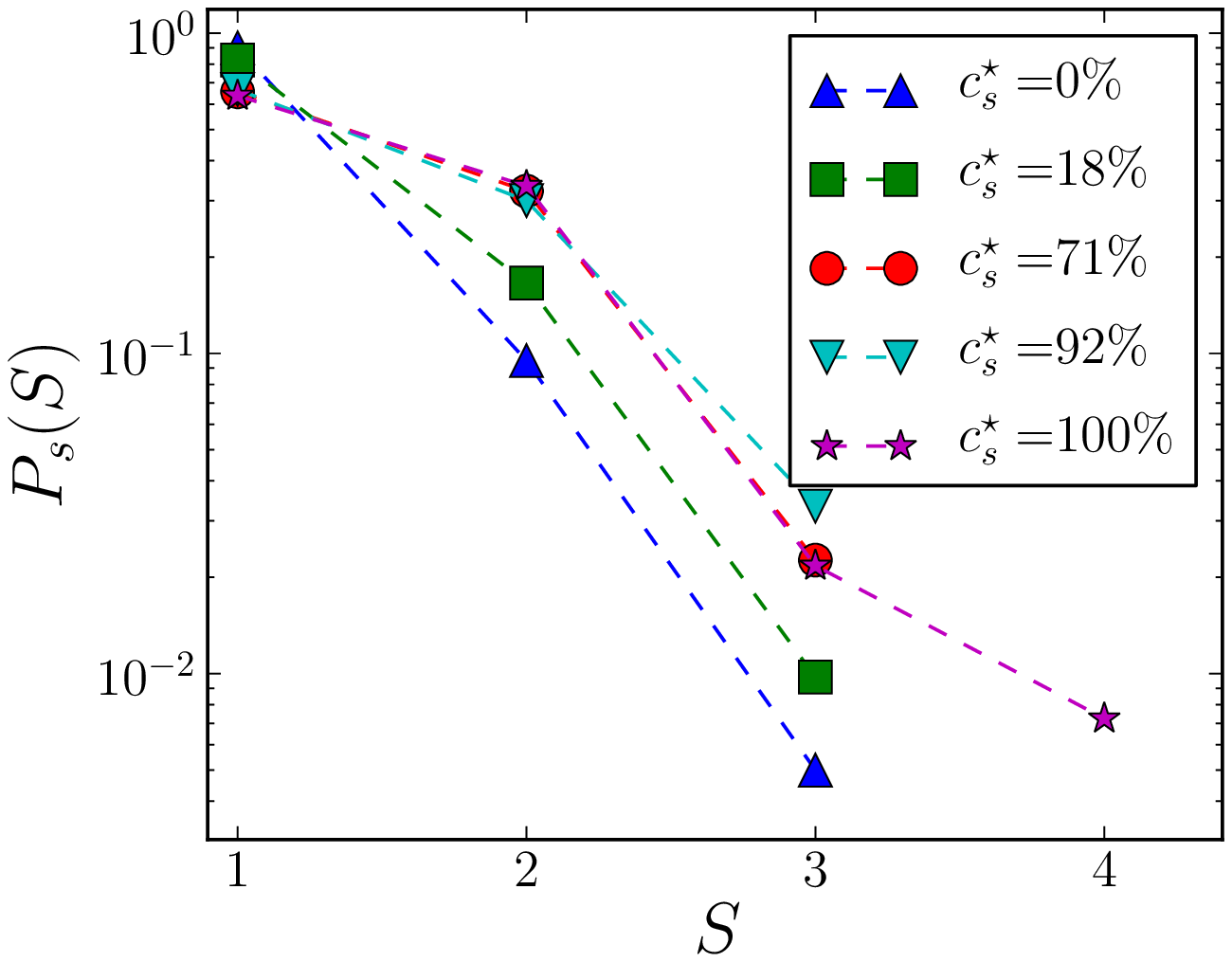}
\caption{Normalised histograms of burst sizes $S$ (i.e., egresses coming in
rapid succession) in the evacuations with hurried walkers.}
\label{fig:burst_sizes}
\end{figure}

The existence of short time lapses $\Delta t$ hints at bursty dynamics. Here,
passages through the door will belong to a burst of escapes if they are
separated by less than 60\% of the mean time lapse of the considered experiment,
viz., $\Delta t< 0.6\,\langle \Delta t \rangle$ (we have checked that
the results are robust to small variations of this criterion). Thus defined,
bursts do not
correspond to intervals of fast flow separated by long clogs (only short-lived
clogs were observed here), but to almost simultaneous escapes. The distribution
$P_s(S)$ of burst sizes is plotted in Fig.~\ref{fig:burst_sizes}. As expected,
egresses of polite crowds consist of isolated escapes, with a small
proportion of bursts of $S=2$ pedestrians or more ($S>2$, with probability 
$P_s(S)<1\%$), whereas bursts of up to $S=4$ pedestrians are observed with more
selfish and competitive crowds. Our data are perfectly compatible with the
(fast) exponential decay of $P_s(S)$ reported in \cite{garcimartin2016flow}, but
are insufficient to 
be assertive in this regard. Incidentally, note that, in all but one experiment
($c_s^\star=71\%$(hurried)), the average fraction of selfish agents in
a
burst increases with the size of the burst. This is particularly noticeable in
the ``placid'' experiments. However, collecting statistics over the entire
evacuation, we have not detected substantial temporal correlations in the
behaviours of successively egressing participants.

\subsubsection*{Alternation between short and long time lapses, generalised
"zipper effect"}
\begin{figure}[ht]
\centering
\includegraphics[width=\linewidth]{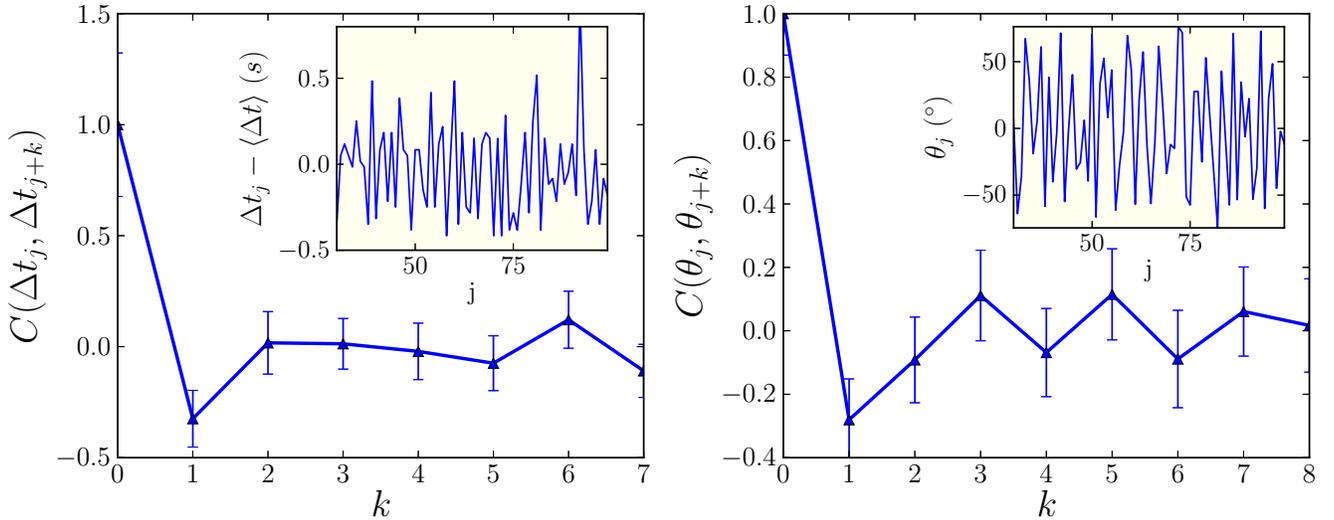}
\caption{Correlations between time lapses $\Delta t_j$ and $\Delta t_{j+k}$,
averaged over $j$ (left) and correlations between the angles of incidence
$\theta$ of a pedestrian and his or her $k$-th successor (right). The error bars
represent 95\% confidence intervals.}
\label{fig:Delta_t_correl}
\end{figure}

Although successions of $S>2$ short time lapses exist (bursts of size $S>2$),
statistically significant anti-correlations between successive $\Delta t$
prevail by far, in
all experiments but two ($c_s^\star=45\%$(placid) and, to a lesser extent,
$c_s^\star=47\%$(placid)). This is apparent in the negative dip in the
correlation functions of successive time lapses $\Delta t_j$ (ordered by the
rank of egress),
\begin{equation*}
C(\Delta t_j, \Delta t_{j+k})\equiv  \frac{ \Big\langle (\Delta t_j - \langle
\Delta t \rangle) (\Delta t_{j+k} - \langle \Delta t \rangle) \Big\rangle}{
\langle (\Delta t_j - \langle \Delta t \rangle)^2 \rangle},
\end{equation*}
where the brackets denote an average over all pedestrians $j$ in the experiment,
an example of which is shown in Fig.~\ref{fig:Delta_t_correl}. This means that,
in general, there is an alternation between shorter time lapses and longer ones.
Such an alternation has already been observed in cooperative pedestrian flows at
the entrance of (long) bottlenecks
\cite{hoogendoorn2003extracting,hoogendoorn2005pedestrian,seyfried2009new}. It
was then ascribed to a ``zipper effect'', whereby pedestrian lanes form in the
bottleneck and need to be intercalated, because the bottleneck width does not
allow pedestrians from different lanes to stand shoulder to shoulder. Within
each lane, the headway that pedestrians maintain with respect to the walker just
in front of them (of order one second in \cite{hoogendoorn2005pedestrian})
imposes a finite minimal distance between the pedestrians. But there is
no such headway between distinct lanes, so that a pedestrian may come close to
contact with a walker
from a neighbouring lane, without 
being able overtake this neighbour because of the constrained lateral space.
Albeit alluring, this scenario does not apply for our experiments, where

strong anti-correlations are also seen in very
competitive and disorderly evacuations, for instance at
$c_s^\star=100\%$(hurried). In the latter, no lanes whatsoever
can form. To
explain the alternation, we believe that a more general mechanism
should be put forward: \emph{If some free
space is available right in front of them}, pedestrians will step forward and
try
to cross the door at (almost) the same time as their predecessor (who comes from
\emph{another} direction). But they will not risk this manoeuvre if there is not
sufficient headway just in front of them, in particular if a competition for the
exit already blocks the door.

To bolster our claim, let us study the pedestrians' angles of incidence $\theta$
into the exit zone, the angle $\theta=0^\circ$ corresponding to normal
(``central'') incidence.  The distributions of $\theta$ are approximately flat
over
an interval of the form $[-\theta_\mathrm{max},\theta_\mathrm{max}]$, where the
maximal angle $\theta_\mathrm{max}$ seems to increase with the competitiveness
of the evacuation (Supplementary Fig.~S4). More relevantly, we remark that the
directions $\theta$
exhibit strong anti-correlations in time, pointing to the prevalence of
alternations between small and large $\theta$. This is striking, for instance,
in the autocorrelation function,
\begin{equation*}
C(\theta_j, \theta_{j+k})\equiv \frac{ \Big\langle (\theta_j - \langle \theta
\rangle)(\theta_{j+k} - \langle \theta \rangle) \Big\rangle}{ \Big\langle
(\theta_j - \langle \theta \rangle)^2 \Big\rangle},
\end{equation*}
plotted in Fig.~\ref{fig:Delta_t_correl}(right) for
$c_s^\star=100\%$(hurried), where the negative dip for successively
egressing
pedestrians ($k=1$) lies more than $4\frac{\sigma}{\sqrt{n}}$ below 0, where
$\sigma$ is the standard deviation of the autocorrelations of $\theta$ (whose
average gives $C$) and $n$ is the number of points used to compute the average.
Moreover, if pedestrians $i$ and $i+1$ egress in fast succession ($\Delta
t<0.6\langle \Delta t \rangle$), on
average their angles of incidence $\theta$ differ more  than if
 their egresses are separated by a longer delay ($\Delta t>1.5\langle
\Delta t \rangle$). This is true in strictly \emph{all} experiments, with the
most marked
contrasts for polite crowds, e.g.,

\begin{equation*}
\langle|\theta_{i+1}-\theta_{i}|\rangle_{fast}=80^\circ \text{
vs. }\langle|\theta_{i+1}-\theta_{i}|\rangle_{slow}=24^\circ \text{ for }
c_s^\star=0\% \text{(hurried)}.
\end{equation*}
It confirms that the short time lapses
correspond to a participant from \emph{another} direction passing through the
door very shortly after the previous one. 
Conversely, the headway left behind a participant coming from the same direction
is reflected by the fact that one systematically has to wait longer (on average
and in median) to see another participant egress from the same direction (within
$25^\circ$) than from a randomly selected direction in the empirical
distribution of $\theta$.
The foregoing discussion implies that time lapses $\Delta t$ longer than the
median value are not due to slower, lazier, or more patient pedestrians, but
result from the microscopic organisation of the constricted flow.

We should mention that alternations between short and long $\Delta t$ are
not seen in the uncontrolled egresses, where people generally
passed through the door one by one, in an orderly and non-competitive way,
sometimes even starting to align beforehand.

\subsection*{Dynamics in the exit zone}
\subsubsection*{Dwell time in the exit zone}
So far we have focused on the time lapses between egresses. But, from the
individual pedestrian's perspective, a possibly more relevant duration is the
time they have to wait in the above-defined exit zone, of area $A$, before they
can egress. Let us call this dwell time (or waiting time) $T_w$. Its mean value
$\langle T_w \rangle$ is related to the mean time lapse via $
\langle T_w \rangle = \langle \rho \rangle A \langle \Delta t \rangle$; the
derivation is presented as Supplementary Information. 

\begin{figure}[ht]
\centering
\includegraphics[width=\linewidth]{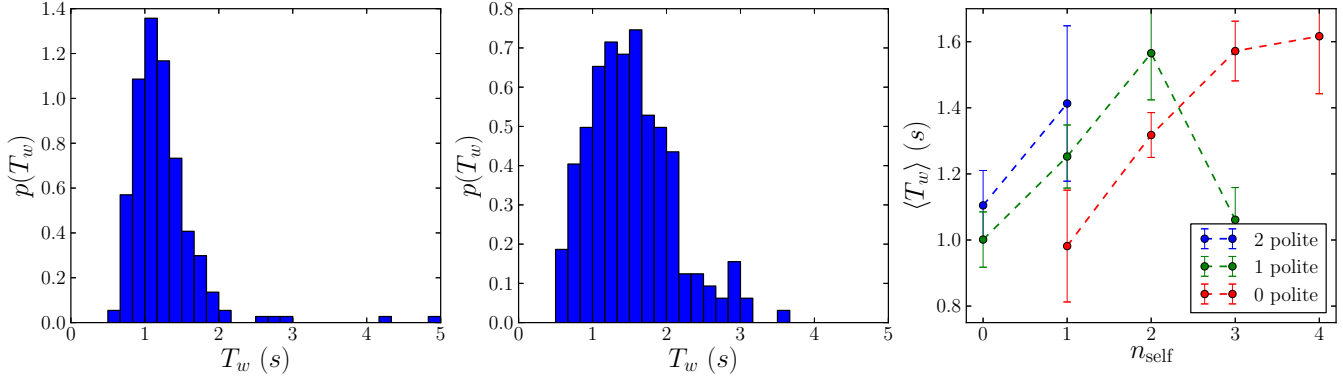}
\caption{Dwell times $T_w$ in the exit zone in evacuations with hurried
walkers.  Normalised histograms of $T_w$ for $c_s^\star=0\%$(hurried)
(left)
and $c_s^\star=100\%$(hurried) (middle). The right panel shows the
average
dwell time for selfish, hurried pedestrians as a function of the number
$n_{\mathrm{self}}$ of selfish participants in the zone, for distinct numbers of
polite participants. The error bars are 95\% confidence intervals.
Note that these averages were computed from subsamples of
$n\approx20-200$ data points, except in the cases of 0 polite agent and 1
selfish one ($n=11$) and 1 polite agent and 3 selfish ones ($n=6$). } 
\label{fig:T_w}
\end{figure}

Sharply peaked distributions $p(T_w)$ correspond to orderly evacuations, with
approximately equal waiting times for all. This case is exemplified in
Fig.~\ref{fig:T_w}(left), for $c_s^\star=0\%$(hurried). On the
contrary,
broad $p(T_w)$ reveal the presence of heterogeneity and/or disorder. For
$c_s^\star=100\%$(hurried), since the crowd is behaviourally
homogeneous, the
breadth of $p(T_w)$ (Fig.~\ref{fig:T_w}(right)) has to be ascribed to disorder.
On the other hand, for inhomogeneous crowds, the overall distribution $p(T_w)$
mingles the two distinct types of pedestrians. 

\begin{figure}[ht]
\centering
\includegraphics[width=290pt]{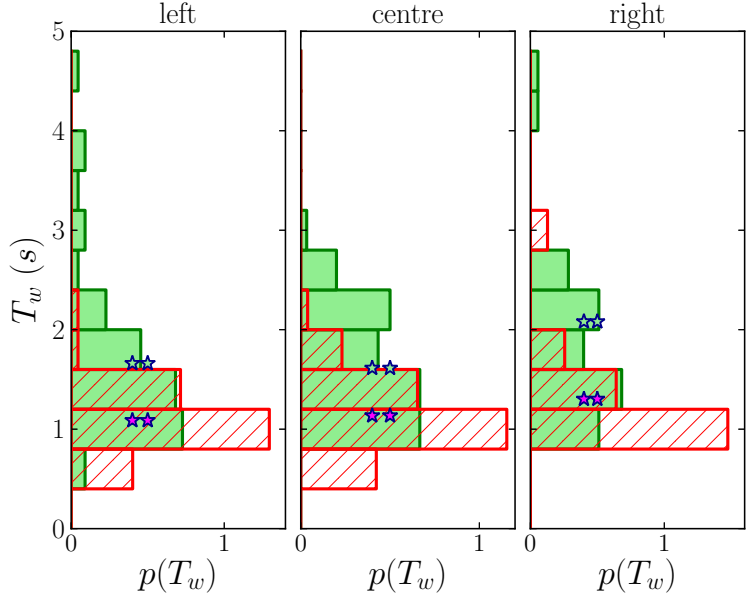}
\caption{Normalised histograms of waiting times in front of the door as a
function of the participant's behaviour (green for a polite pedestrian, red for
a selfish one) and the direction (left, centre or right) in which they entered
the zone, in an evacuation with $c_s^\star=46\%$. The stars indicate the mean
value in each case.}
\label{fig:T_w_dist_direc}
\end{figure}

To refine the analysis, we study the histograms of $T_w$ as a function of the
pedestrian's behaviour, namely, whether they are polite agents (PolA) or selfish
moving agents (SelMA), and their direction of incidence into the exit zone: from
the left ($\theta<60^\circ$), from the centre ($60^\circ\leqslant \theta <
120^\circ$), or from the right ($\theta \geqslant 120^\circ$). An example of
such a histogram is shown in Fig.~\ref{fig:T_w_dist_direc}. Beyond the
statistical noise, we see that, in all cases, the distributions are peaked at a
value around 1s and, as expected, PolA's distributions are shifted to larger
$T_w$, and stretched, compared to SelMA's. For either type of behaviour, the
distributions look relatively similar for all incident directions \footnote{We
observed a surprising asymmetry between right and left, in some cases, e.g.,
$c_s^\star=71\%$(hurried), where the average waiting time differs by
more
than $2.6\frac{\sigma}{\sqrt{n}}$. This could be due to statistical noise,
individual preferences for one side, or a slight, unnoticed asymmetry in the
setup.}, with
a noticeable exception: PolAs display more frequent outliers at large $T_w$ if
they come from the sides than if they come from the centre. Indeed, PolA tends
to be swept along by the pedestrian stream if she comes from the centre, whereas
she can be blocked at the sides of the door otherwise. There is no such a marked
effect for SelMA, probably because the latter tries her hardest, and manages, to
egress rapidly in any event.

How do the $T_w$ depend on the population in the exit zone (averaged over the
pedestrian's dwell time)? Figure~\ref{fig:T_w}(right) evinces that, in the
 experiments with hurry, SelMA's mean waiting time very generally grows
with the
number of selfish agents in the zone, for any fixed number of polite ones, and
also increases when there are more polite participants, for a fixed number of
selfish ones. These features hold true in the placid experiments (Supplementary
Fig.~S5), with the exception that $\langle T_w \rangle $ seems to shorten in the
competitive situations where 3 participants are simultaneously in the zone,
perhaps due to a sampling bias. On the whole, these observations also apply for
PolA's waiting times, but the picture is less clear. The positive correlation
between the waiting time and the occupancy of the zone is rather intuitive, but
it is not trivial, for at least two reasons. First, it implies that
statistically the influence of the global evacuation competitiveness on $T_w$ is
outshone by the effect of some local parameters, more precisely, the exit zone
occupancy. Secondly, escapes come in faster succession (shorter $\Delta t$) if
the zone is more crowded.

\subsubsection*{Priority rules in the zone}
The aforementioned correlations are only valid in a statistical sense. We have
not found any law that can accurately predict $T_w$ as a function of the local
parameters. Nevertheless, can some generally valid ``priority rules'' for the
flow be unveiled? Let us study to whom an
agent of a given type yields, that is to say, who entered
the zone later than this
agent but egressed before. By listing these pedestrians, we find that overall
SelMA yields to fewer
pedestrians than PolA and that, for both types, it is more common to yield to
selfish pedestrians than to polite ones, which is not so surprising. It should
however be stressed that this tendency is not systematic: polite agents
(possibly carried along by the pedestrian flow) also overtake their selfish
counterparts from time to time. On the other hand, in virtually no instance does
SelMA yield to a polite pedestrian coming from the same direction (left, centre
or right). This rule only suffers two exceptions over the whole set of
experiments.

\subsubsection*{Disorder, pressure, and pedestrian ``vortices''}
Clearly, these ``priority rules'' are limited in scope and cannot tell us how
orderly the flow is. To quantify disorder, we choose to compute the average
difference $D^1$ between the order of entrance $k_{in}(i)$ of pedestrian
\emph{i} in the zone and their order of egress $k_{out}(i)$, viz., $D^1 \equiv
\frac{1}{N} \sum_{i=1}^{N} |k_{out}(i)-k_{in}(i)|$. The disorder estimator $D^1$
increases with $\rho$ (see Supplementary Fig.~S6), but also with $c_s^\star$;
which of these two parameters is best correlated with $D^1$ is not obvious. But
it is noteworthy that the crowd at $c_s^\star=92\%$ has a larger $D^1$ value
than that at $c_s^\star=100\%$, which points to the enhancement of disorder due
to behavioural heterogeneity.

Besides disorder, another practically relevant feature of the evacuation is the
pressure exerted on the participants. Indeed, suffocation and
compressive asphyxia, originating in the compression of the lungs, have been
reported as a cause of death in major crowd disasters \cite{helbing2012crowd}.
Accordingly, after each experiment, we asked a dozen randomly picked
participants to rate the level of mechanical pressure that they experienced
during the evacuation from 1 to 10, for want of more objective measurements of
pressure. Admittedly, this is a subjective evaluation by the
participants, which will yield mostly qualitative results. Still, the perceived
pressure follows a clear trend, more precisely, a
monotonic increase with the density $\rho$ in the exit zone (and
thus also with $c_s^\star$), as shown in Supplementary
Fig.~S7).
Therefore, competitive evacuations, associated with larger densities at the
door, involve more disorder and higher pressure than their more cooperative
counterparts. The most competitive ones even feature signs of incipient
turbulence, in the form of ``vortices'' at the exit, i.e., participants spinning
around a full 360$^\circ$ when egressing (see Table~\ref{tab:global} and videos
in the Supplemental Materials).

\section*{Summary and outlook}
In summary, we have performed pedestrian flow experiments through a
72-cm-wide door,
in which a (variable) fraction $c_s^\star$ of the participants were asked to
behave selfishly, while the rest behaved politely. Irrespective of these
behaviours, a first series of experiments was conducted with placid walkers and
a second one with hurried walkers. By re-injecting egressing
participants
into the room, we managed to improve the statistics. Furthermore, despite
instantaneous fluctuations, the pedestrian flow was found to be quasi-stationary
at the macroscopic scale, unlike in other experiments
\cite{seyfried2009new,garcimartin2016flow}.

The flow rate $J$ gets higher for larger $c_s^\star$ and with more
hurried
walkers. The absence of a `faster-is-slower' effect does not contradict its
possible occurrence
\cite{Garcimartin2014experimental,Pastor2015experimental} when
one considers
crowds above a competitiveness threshold. 

Regardless of the behaviours, the flow rate and other flow properties such as
the disorder in the passages and the pressure perceived by the participants
exhibit a simple dependence on the density $\rho$ in the exit zone. Thus, our
results suggest that in a coarse macroscopic approach,  and for a given
composition of the crowd, the behavioural aspects can be left aside in favour of
the density $\rho$; this confirms the key role played by the latter in
determining the flow rate \cite{seyfried2009new}. These variations with
$\rho$ suggest corrections to simple traffic theories based on the
assumption that at
a bottleneck the system adopts the local parameters that maximise the flow
rate, if the inflow is sufficiently large. Instead, our findings indicate that
the selected density at the bottleneck is controlled by the pedestrians'
behaviours.

Somewhat surprisingly, $J$ was
observed to grow monotonically with $\rho$ up to close-packing ($\rho \approx
9\,\mathrm{m^{-2}}$), despite the jams predicted by most
fundamental diagrams in the literature at these densities
\cite{predtechenskii1978planning,weidmann1993transporttechnik,dinenno2008sfpe,
flotterod2015bidirectional}.
In fact, we believe that at high densities the global flow rate will strongly
depend on other parameters, in particular, the pressure in the crowd, thus, to
what extent people are pushing their neighbours. Here, the participants were not
allowed to 
push. Yet, the evacuation dynamics clearly became more disorderly with
increasing $c_s^\star$ and increasing hurry. This disorder was notably
reflected in the distribution $p(\Delta t)$ of time lapses $\Delta t$ between
successive escapes: $p(\Delta t)$ displays a relatively sharp peak at $\Delta
t^\star$ of order $1\,\mathrm{s}$ for polite crowds. In more competitive
evacuations the low-$\Delta t$ region  gets populated; bursts of
quasi-simultaneous escapes occur and seem to be exponentially distributed in
size. In the most competitive evacuations, some signs of incipient flow
turbulence were even detected, e.g., pedestrian ``vortices''. To what
extent these features can be extrapolated to extreme conditions of emergency
still needs to be ascertained.

Shorter and longer time lapses $\Delta t$ were found to alternate between
successive escapes in almost all experiments. We explained this in terms of a
generalised zipper effect, whereby pedestrians strive to keep a finite headway
behind an agent walking in the same direction, while coming close to contact
with those from another direction. This idea is supported by the marked
anti-correlations in the angles of incidence at the door.

Finally, we investigated the pedestrians' dwell time $T_w$ in the exit zone. The
mean dwell time clearly increases with the occupancy of the zone, especially for
selfish agents. Not surprisingly, $T_w$ is shorter and more narrowly distributed
for selfish agents. Also, polite pedestrians coming from the sides display more
outliers at large $T_w$.

Several of the effects observed in our ``microscopic'' analysis of the
evacuation
dynamics are rather intuitive. But their quantitative characterisation opens the
door (so to speak) to more thorough tests of pedestrian flow models, as in
Ref.~\cite{robin2009specification}, and, accordingly,
more precise understanding of the process. Our study may also prompt the idea
that the effect of the pedestrians' behaviours on the evacuation dynamics is
amenable to a simple description.

\section*{Acknowledgements}
The authors are grateful to the Grupo de Higiene y Seguridad of CAB for their
help in devising safe evacuation
experiments, to all members of the FiEstIn group, in particular to Guillermo
Abramson and Dami\'an
Zanette, to Alejandro Kolton and Pablo Gleiser for lending us their cameras, and
to all the voluntary participants. A.N. thanks Iker Zuriguel for a useful
discussion.
This work received partial funding from CONICET (under Grant No. PIP
11220110100310) and CNEA, both Argentinian agencies.

\section*{Author contributions statement}
A.N., S.B., and M.K. conceived and conducted the experiments. A.N. analysed the
results and wrote the paper. All authors discussed the analysis and reviewed the
manuscript. (S.B. and M.K. contributed equally to this work.)

\section*{Additional information}
The authors declare no competing financial interests.

\end{document}